\documentclass[prb,12pt]{revtex4}

\usepackage[english]{babel}
\usepackage[intlimits]{amsmath}
\usepackage[dvips]{graphicx}

\begin{document}

\begin{titlepage}
\
\title{Simple Growth Models of Rigid Multifilament Biopolymers}
\author{Evgeny B. Stukalin and Anatoly B. Kolomeisky}

\affiliation{Department of Chemistry, Rice University, Houston, TX 77005 USA}

\begin{abstract}

{The growth dynamics of rigid biopolymers, consisting of $N$ parallel protofilaments, is investigated theoretically using simple approximate models. In our approach, the structure of a polymer's growing end and lateral interactions between protofilaments are explicitly taken into account, and  it is argued that only few  conformations are important for biopolymer's growth. As a result, exact analytic expressions for growth velocity and dispersion are obtained for {\it any} number of protofilaments and arbitrary geometry of the growing end of the biopolymer. Our theoretical predictions are compared with a full description of biopolymer growth dynamics for the simplest $N=2$ model. It is found that the results from the approximate theory are approaching the exact ones for large lateral interactions between the protofilaments. Our theory is also  applied  to analyze the experimental data on the growth of microtubules.}

\end{abstract}

\maketitle

\end{titlepage}

\section {Introduction}

Rigid biopolymers such as microtubules, actin filaments and intermediate filaments are major components of cytoskeleton and cellular environment.  They play a fundamental role in biological systems  by supporting cellular transport, cell motility and reproduction.\cite{bray_book, howard_book,desai97,howard03} Many cellular processes essential for life are driven by polymerization/depolymerization  dynamics of these biopolymers. Therefore, a full theoretical description of growth processes is clearly needed in order to understand mechanisms and principles of cell functioning. 

Microtubules are  rigid, hollow tubular biopolymers made of parallel protofilaments arranged in circular array.\cite{bray_book, howard_book,desai97}  Each protofilament is a linear polymer chain consisting of alternating $\alpha$- and $\beta$-tubulin subunits. The number of protofilaments varies between 10 and 15 for microtubules from different species, but typically most of them have $N=13$ protofilaments. Lattice structure of microtubules can be viewed as  3 parallel helices, the so-called 3-start helix.\cite{bray_book, howard_book,desai97} It also shows a discontinuity or seam, and a functional role of this lattice feature is unknown. The dynamics of microtubules features an unusual phenomenon of alternating between the growing and shrinking phases which is termed dynamic instability. Actin filaments are another example of rigid biopolymers. They can be described as two-stranded helices in which each actin monomer contacts four other monomers, with the strongest interaction along the strands.\cite{bray_book}.

Recent experiments\cite{dogterom97,dogterom03,mcgrath03} have provided extensive measurements of  growth dynamics of actin filaments and microtubules under the effect of external forces. A number of theoretical models aimed to describe the dynamics of growing rigid biopolymers  have been proposed.\cite{peskin93,mogilner99,van Doorn00,carlsson00,kolomeisky01,dickinson02,freed02}  Several studies  utilized  polymerization ratchet models, which assume that thermal fluctuations at the tip of growing biopolymers control the growth dynamics.\cite{peskin93,mogilner99,van Doorn00,carlsson00} In these models a rigid biopolymer is viewed as consisting of $N$ independent and {\it not-}interacting parallel protofilaments. A different approach is to describe the biopolymer's growth dynamics phenomenologically, by considering the overall polymerization and depolymerization processes and neglecting the microscopic details.\cite{ kolomeisky01,freed02}  Although current theoretical models provide a reasonable description of many aspects of microtubules and actin filaments growth,  there are still many open questions.\cite{kolomeisky01} The general deficiency of current theoretical approaches is the fact that they mainly ignore the microscopic structure and geometrical properties of a biopolymer's lattice, and they also neglect  the lateral interactions between protofilaments.

The purpose of this paper is to investigate the growth dynamics of rigid multi-filament biopolymer by taking into account a complex structure of polymer's growing end, its geometrical properties and interactions between parallel protofilaments. We show that at realistic conditions only few polymer configurations contribute into the overall dynamics. It allows us to develop a simple approximate one-layer model for rigid biopolymers growth, for which we obtain explicit expressions for velocity and dispersion for {\it any} number of protofilaments and for arbitrary geometry of the polymer end. The essential issue of this approximate  treatment proves out to be an evaluation of analytical expressions for the asymptotic (long-time) mean growth velocity, {\it V}, 
\begin{equation}
\langle x(t)\rangle \approx Vt,
\end{equation}
and for  dispersion (or effective diffusion constant), {\it D},
\begin{equation}
D \approx \frac{1}{2}\frac{d}{dt}\left[ \langle x^2(t) \rangle- \langle x(t) \rangle^{2} \right].
\end{equation}
Here, {\it x(t)} stands for a coordinate of the biopolymer's tip at time {\it t} that grows linearly with time at stationary-state limit. Dispersion {\it D} is a natural measure of fluctuations of growth dynamics. Note, that although in this paper we aim to describe the microtubule dynamics, it could also be used to analyze the growth of other rigid biopolymers, such as actin filaments.   

The paper is organized as follows:  the approximate model of the growth of biopolymer consisting of $N$ protofilaments is presented in Sec. II, while  in Sec. III the full description of growth model with $N=2$ protofilaments is given and compared with the approximate approach. Sec. IV compares the predictions of the approximate model with the exact growth model for $N=2$ case, and  in Sec. V  our theory is used to describe the real experimental data on the growth of microtubules. Sec. VI summarizes all results and concludes our paper.

\section {One-Layer Model}

Consider a growing rigid polymer consisting of N protofilaments as shown in Fig. 1. The building block of such polymers is a  monomer subunit of  length $d$. For microtubules this is an $\alpha$-$\beta$-tubulin heterodimer, which has the length of 8.2 nm.\cite{bray_book, howard_book,desai97}  The protofilaments are parallel to each other but shifted by arbitrary distances. There are chemical interactions between protofilaments  that bind them together in a polymer lattice. Protofilaments are labeled in a such  way that the seam in the polymer lattice is always between protofilaments 1 and $N$. There is an infinite number of possible polymer configurations, which can be described by a set of $N$ numbers, $\{a_{1}, a_{2}, \ldots, a_{N}\}$, where $a_{j}$ is a coordinate of the tip of the protofilament $j$. For labeling different configurations, let us choose a moving coordinate system  where the origin is always at the tip of the leading protofilament, i.e., $a_{j}=0$ if the protofilament $j$ is the leading one: see Fig. 1.

When a free  monomer  approaches the growing biopolymer molecule, it can attach to any of $N$ protofilaments. Define $u_{j}$ ($w_{j}$) as a  rate of attachment (detachment) to the protofilament $j$. These rates are related by the following thermodynamic expression,
\begin{equation}
\frac{u_{j}}{w_{j}}=\exp [-(g_{v}+g_{h}+g_{im})/k_{B}T],
\end{equation} 
where $g_{v}$ is a  bond energy due to head-to-tail binding (vertical), $g_{h}$ is the  energy due to lateral (horizontal) interaction, and $g_{im}$ is the free energy of immobilizing the free monomer into the rigid lattice:\cite{erickson89} see Fig. 1. For binding to any protofilament the value of longitudinal energy  $g_{v}$  is the same, however, contributions to the horizontal energy $g_{h}$ might be different because the local environment of each protofilament tip is  different. For polymerization/depolymerization rates we can easily write
\begin{equation}
u_{j} \propto \exp [-\theta(g_{v}+g_{h}+g_{im})/k_{B}T], \quad w_{j} \propto \exp [-(\theta-1)(g_{v}+g_{h}+g_{im})/k_{B}T];
\end{equation}  
where the coefficient $\theta$,  $0<\theta<1$, reflects the value of the activation barrier for the process of monomer binding. The exact values of the lateral interaction energy $g_{h}$ for microtubules are unknown, but there is an estimate based on computer simulations of a stochastic model of microtubule assembly dynamics, which gives  $g_{h}$ approximately between -3 and -6 $k_{B}T$. \cite{vanburen02} It is obvious that the monomer could attach  much faster to some protofilaments when more lateral bonds are made. That leads to observation that the growing biopolymer can be found preferentially in some particular configurations. This is the basis for our approximate model. 

In our model we assume that the growing biopolymer can only be found in configurations with distances from all protofilament tips less than $d$, i.e., $a_{j}<d$ for all $j$, as shown in Fig. 2. It means that all protofilament ends are within one monomer distance from the leading protofilament. There are $N$ such configurations, because each protofilament can be the leading one only once. This is a {\it one-layer} model of rigid biopolymer growth.

Consider the dynamics of  biopolymer growth  in the one-layer model. For the biopolymer in the configuration $j$, i.e. when the tip of the protofilament $j$ is at the origin, the incoming monomer can only attach to one protofilament, which is the furthest away, and the polymer configuration transformed into another one. The whole new layer of the monomers of length $d$ is added to the biopolymer only when the system goes sequentially  through all allowed $N$ configurations, and we return to the same configuration $j$. Thus, the growing biopolymer advances from the given configuration  to the same one, but only  shifted by distance $d$, through the sequence of $N$ states. Then the process repeats again and again. From mathematical point of view, this dynamic  description can be mapped into the motion of a single particle on  periodic one-dimensional lattices.\cite{derrida83} Recently, similar approach has been used successfully to describe the dynamics of motor proteins.\cite{FK01,KF03} 

This mapping allows us to obtain  exact and explicit expressions for  the stationary-state growth velocity and the dispersion, or  effective diffusion constant, for the rigid biopolymer with {\it any} number of protofilaments and for any arbitrary set of shifts $\{a_{1},a_{2},\cdots,a_{N}\}$ in terms of rate constants $\{u_{j},w_{j}\}$. The equation  for the steady mean velocity of growth is given by\cite{derrida83,FK01,KF03}
\begin{equation}\label{velocity}
V=\frac{d}{R_{N}} \left( 1-\prod_{j=1}^{N} \frac{w_{j}}{u_{j}} \right),
\end{equation}
where $d$ is the size of the monomer subunit, while
\begin{equation}\label{velocity.add}
R_{N}=\sum_{j=1}^{N} r_{j}, \quad r_{j}=\frac{1}{u_{j}} \left(1+ \sum_{k=1}^{N-1} \prod_{i=1}^{k} \frac{w_{j+i}}{u_{j+i}} \right).
\end{equation}
Here we also used the periodicity of the one-layer model, i.e., $u_{j \pm N}=u_{j}$ and  $w_{j \pm N}=w_{j}$.

The expression for the dispersion is similar but more complex:\cite{derrida83,FK01,KF03}
\begin{equation}\label{dispersion}
D=\frac{d}{N} \left\{\frac{VS_{N}+dU_{N}}{R_{N}^{2}}-\frac{1}{2}(N+2)V  \right \},
\end{equation}
where the auxiliary functions are given by
\begin{equation}
S_{N}=\sum_{j=1}^{N}s_{j} \sum_{i=1}^{N}ir_{i+j}, \quad U_{N}=\sum_{j=1}^{N} u_{j}r_{j}s_{j},
\end{equation}
with supplementary coefficients
\begin{equation}
s_{j}=\frac{1}{u_{j}} \left(1+ \sum_{k=1}^{N-1} \prod_{i=1}^{k} \frac{w_{j+1-i}}{u_{j-i}} \right).
\end{equation}

In the simplest case of the growth of the biopolymer with $N=2$ protofilaments these expressions are reduced  to
\begin{equation}\label{vel.N=2}
V=d\frac{u_{1}u_{2}-w_{1}w_{2}}{u_{1}+u_{2}+w_{1}+w_{2}},
\end{equation}
for the mean growth velocity, while for the dispersion it yields
\begin{equation}
D=\frac{d^{2}}{2} \frac{u_{1}u_{2}+w_{1}w_{2}-2(V/d)^{2}}{u_{1}+u_{2}+w_{1}+w_{2}}.
\end{equation}

\subsection{Effect of External Loads}

Forces produced by growing microtubules are crucial for understanding mechanisms of cellular motility and cellular transport.\cite{bray_book, howard_book,desai97,howard03} To investigate force production in experiments the  growing microtubules are  put under external loads which slow down the growth.\cite{dogterom97,dogterom03} The experimentally used external loads are hard walls and/or optical trap systems. These studies provide a valuable information on biopolymer growth mechanisms and cellular motility, and any theoretical description should account for the effect of external forces. In the one-layer model this  can be easily done.

Consider a microtubule which is growing against an external  force $F$. This force  acts locally only on the leading protofilament. An example will be a hard wall positioned at the tip of the leading protofilament. When a  monomer is attaching to the protofilament $j$, the microtubule produces the work and it is equal to $F(d-a_{j})$. Then the rates of polymerization and depolymerization should be modified as follows, 
\begin{eqnarray}\label{vel.force}
u_{j}(F)= u_{j}(0)exp[-\theta_{j}^{+} F(d-a_{j})/k_{B}T], \nonumber \\
w_{j}(F)=w_{j}(0)exp[+\theta_{j}^{-}F(d-a_{j})/k_{B}T]
\end{eqnarray}
where $d-a_{j}$ is the microtubule length change for monomer binding to the protofilament $j$,  and $\theta_{j}^{+}$ and $\theta_{j}^{-}$ are the {\it load-distribution factors}. These factors reflect how the external force affects the activation energy for attachment and detachment processes of the monomer subunit.\cite{FK01,KF03} Also, the load-distribution factors may have different signs, but we certainly expect that the overall factor
\begin{equation}
 \theta= \sum_{j=1}^{N}(\theta_{j}^{+}+\theta_{j}^{-}),
\end{equation}
to be positive, implying an opposition to growth of the biopolymer molecule. The force-dependent rate constants can be substituted then in Eqs. (\ref{velocity}) and (\ref{dispersion}) to obtain the load-dependent explicit expressions for the velocity and dispersion, and thus providing a full description of biopolymer growth under external forces (see also Fig. 6).

If the external force opposes to polymerization, then there is a special value of force, termed stalling force, $F_{S}$, at which the force-dependent mean growth velocity is approaching zero. This is an important characteristic of real  biopolymers.\cite{howard_book} In our approximate model we can easily calculate the stalling force,\begin{equation}\label{stallforce}
 F_{S}=\frac{k_{B}T}{d} \ln \frac{\prod_{j=1}^{N} u_{j}}{\prod_{j=1}^{N} w_{j}}.
\end{equation}
Similar expressions have been used successfully for the description of motor proteins dynamics.\cite{FK01,KF03}

\subsection{Comparison with Phenomenological Description}

It is very interesting to compare our approximate approach with phenomenological models which dominate in the field of microtubule dynamics. According to a  phenomenological description, the  growth rate is determined by the balance between polymerization and depolymerization processes,
\begin{equation}\label{vel.phenomen}
V=\frac{d}{N}(k_{on}c-k_{off}),
\end{equation}
where $c$ is the concentration of free tubulin-GTP subunits, and $k_{on}$ and $k_{off}$ are average rate constants for polymerization and depolymerization, respectively. This simple picture suggests that there is a linear dependence of microtubule growth on tubulin concentration, at least, in the regime where the dynamic instability can be neglected.

In the one-layer model, binding rate constants are also proportional to the concentration of free monomers, $u_{j}=k_{j} c$. However, the linear dependence of the mean growth velocity is only valid at large concentrations of monomers, while at low concentrations the behavior is  different. The mean growth velocity has a power law dependence on the concentration at this regime. It can be easily seen by analyzing Eq. (\ref{vel.N=2}) for the  case of the biopolymer growth with $N=2$ biopolymers.  At large $c$, one can obtain $V \propto \frac{k_{1}k_{2}}{k_{1}+k_{2}}c$, while at low $c$ the mean growth velocity has a quadratic dependence. Similar deviations are also found for dispersion as a function of the free monomers concentration. We expect that in general $N$ case the behavior is analogous.

The observation that the biopolymer growth velocity may deviate from the simple linear dependence given in Eq. (\ref{vel.phenomen}) has  serious consequences since all experimental measurements have been based on this phenomenological picture. It means that the values of measured rate constants $k_{on}$ and $k_{off}$ may be different for different tubulin concentrations. This could explain a large concentration variability in growth rates of microtubules.\cite{desai97,pedigo02}  This question requires a careful experimental investigation.

\section{Growth Dynamics for Biopolymers with $N=2$ Protofilaments}

How well the one-layer model approximates the full growth dynamics of rigid biopolymers  is an open question. However, we can answer this question in the simplest non-trivial  case of the biopolymer with $N=2$ protofilaments, where an exact solution can be found for the full dynamical description. A comparison between the approximate and the complete dynamical models provides an important physical insight into the mechanism of rigid biopolymer growth, and allows us to understand the applicability of the one-layer model.

When the monomer is binding or dissociating from the growing biopolymer with $N=2$ protofilaments, there are four different situations: see Fig. 3. The monomer can attach with the rate $u_{1}$ to the protofilament 1 of the configuration $\{ a,0 \}$, where $a$, $0<a<d$, is the shift between the protofilaments, or it can dissociate with the rate $w_{1}$ from the protofilament 1 of the configuration $\{ 0,d-a \}$, as shown in Fig. 3a. These rate constants are related through the thermodynamic expression
\begin{equation}
\frac{u_{1}}{w_{1}}=\exp \left(-\frac{2 g_{h} a/d +g_{v}+g_{im}}{k_{B}T} \right),
\end{equation}
where $g_{h}$ is the the energy of lateral (horizontal) interaction per one subunit,  $g_{v}$ is the longitudinal (vertical) bond energy, and $g_{im}$ is the entropic term which describes the energy of immobilization. Then the factor $2a/d$ gives the fraction of the lateral bond  created after the monomer binded  to the biopolymer. 

Similarly, the monomer can attach with the rate $u_{2}$ to the protofilament 2 of the configuration $\{ 0,d-a \}$, or  it can dissociate with the rate $w_{2}$ from the protofilament 2 of the configuration $\{ a,0 \}$, as shown in Fig. 3b. The relation for these rates is given by
\begin{equation}
\frac{u_{2}}{w_{2}}=\exp \left(-\frac{2 g_{h}(1- a/d) +g_{v}+g_{im}}{k_{B}T} \right).
\end{equation}

Rate constants $u_{3}$ and $w_{3}$ describe the polymerization/depolymerization processes, which do not change the length of the polymer, as illustrated in Fig 3c. Meanwhile, the attachment and detachment rates from the leading protofilament for all possible configurations are given by $u$ and $w$, respectively, as shown in Fig. 3d. These rate constants can be described by the following expressions,
\begin{equation}\label{u/w}
\frac{u_{3}}{w_{3}}=\exp \left(-\frac{2 g_{h}  +g_{v}+g_{im}}{k_{B}T} \right), \quad \mbox{and} \quad \frac{u}{w}=\exp \left(-\frac{g_{v}+g_{im}}{k_{B}T} \right).
\end{equation}
For convenience, let us define $\gamma$ as a new energy parameter that specifies lateral interactions,
\begin{equation}
\gamma=exp\left(-g_{h}/k_{B}T\right).
\end{equation}
Then,  using Eq. (\ref{u/w}), we obtain a set of simpler relations between the rate constants, namely,
\begin{eqnarray}\label{thermo}
u_{1}/w_{1}&=&u/w\:\gamma^{2a/d} \nonumber \\
u_{2}/w_{2}&=&u/w\:\gamma^{2(1-a/d)}\nonumber \\
u_{3}/w_{3}&=&u/w\:\gamma^{2}
\end{eqnarray} 

Let us define $P( a+id, 0)$ as a probability to find the system in a configuration, where the protofilament 2 is the leading one and the tip of the first protofilament is at distance $a+id,\: (i=0,1,...)$. Similarly, we define $P(0,d-a+id)$ as a probability to find  a configuration with the tip of the second protofilament at distance $d-a+id,\: (i=0,1,...)$ from the tip of the first protofilament, which is now the leading one. The overall kinetic scheme of the system is shown in Fig. 4. Because of the symmetry, at stationary state the overall flux through the system is equal to zero. Then  the  following relations for probabilities are valid,
\begin{equation}\label{vel1}
P(a+id,0)=\left(\frac{u+w_{3}}{u_{3}+w}\right)^{i}P(a,0);
\end{equation}
\begin{equation}\label{vel2}
P(0,d-a+id)=\left(\frac{u+w_{3}}{u_{3}+w}\right)^{i}P(0,d-a);
\end{equation}
\begin{equation}
P(0,d-a)=\frac{u_{1}+w_{2}}{u_{2}+w_{1}} P(a,0);
\end{equation}
with a normalization condition  
\begin{equation}
\sum_{i=0}^{\infty} \left[ P(a+id,0)+P(0,d-a+id) \right]=1
\end{equation}

It can be argued that the parameter $\frac{u+w_{3}}{u_{3}+w} \equiv \beta$ is less than 1. This follows from the observation that  $u_{3}>u$ and $w_{3}<w$, i.e., the monomer binds faster and stronger to interior of the polymer than to the tip of the leading protofilament. After summation of geometrical series in Eqs. (\ref{vel1}) and (\ref{vel2}), we obtain   
\begin{equation} \label{config}
P(a,0)=\frac {1-\beta}{1+\alpha} \quad \mbox{and} \quad P(0,d-a)=\frac {\alpha(1-\beta)}{1+\alpha},
\end{equation}
where we also defined
\begin{equation}\
\alpha=\frac{u_{1}+w_{2}}{u_{2}+w_{1}}.
\end{equation}
It is interesting to note that $1-P(a,0)-P(0,d-a)$ gives the fraction of configurations that are neglected in our approximate one-layer model. Using Eqs. (\ref{config}) , the simple calculation for this quantity yields  $\beta$.

The mean growth velocity  can be written as a sum of several terms, namely,
\begin{eqnarray}\label{vel.general}
V&=&du+(d-a)u_{1}P(a,0)+au_{2}P(0,d-a) \nonumber \\
& & -dw(1-P(a,0)-P(0,d-a))-(d-a)w_{1}P(0,d-a)-aw_{2}P(a,0).
\end{eqnarray} 
The first positive term corresponds to adding a monomer to the leading protofilament and increasing the length of the polymer by $d$. This may take place at any polymer configuration. The second and third positive terms reflect the addition to ``one-layer'' configurations, i.e., where tips of protofilaments are at the distances less than $d$ from each other. Similarly, the first negative term gives the contribution from dissociation of the monomer from the leading protofilament. It shortens the polymer by distance $d$ for all configurations except ``one-layer'' configurations. The last two terms represent negative contributions to the mean growth velocity from the ``one-layer'' configurations.

Substituting expressions (\ref{config}) into the equation (\ref{vel.general}) yields the final expression for the mean growth velocity of the rigid biopolymer with $N=2$ protofilaments, 
\begin{equation}
V=d \left\{ u-w\beta+(1-\beta)\:\frac{u_{1}u_{2}-w_{1}w_{2}}{u_{1}+u_{2}+w_{1}+w_{2}} \right\}.
\end{equation}
Similar calculations can be performed  for the dispersion, or effective diffusion constant.

In the ``one-layer'' model for biopolymer growth with $N=2$ protofilaments, there are only two configurations, $(a,0)$ and $(0,d-a)$. The system can transfer from the configuration $(a,0)$ to the configuration $(0,d-a)$ with the rate $u_{1}+w_{2}$, or it can go backward with the rate $u_{2}+w_{1}$. At $t \rightarrow \infty$, the probabilities to find the system in different configurations can be easily calculated to give
\begin{equation}
P(a,0)=\frac {1}{1+\alpha}, \quad \mbox{and} \quad P(0,d-a)=\frac {\alpha}{1+\alpha}.
\end{equation}
These equations can also be obtained from the general expressions (\ref{config})  for the full dynamic case when $\beta=0$, i.e. when we neglect all configurations except ``one-layer'' configurations. Then the  mean growth velocity can be easily calculated as follows,
\begin{equation}
V=d\:\frac{u_{1}u_{2}-w_{1}w_{2}}{u_{1}+u_{2}+w_{1}+w_{2}}
\end{equation}

Now, we can compare the predictions of the full dynamics  and ``one-layer''  $N=2$-model for different values of the parameter $\gamma$, which reflects the energy of lateral interactions between the protofilaments. We can rewrite the thermodynamics expressions (\ref{thermo}) in the following form,
\begin{eqnarray}
u_{1}=u \gamma^{f_{1}+a/d}, & u_{2}=u \gamma^{f_{2}+(1-a/d)}, & u_{3}=u \gamma^{f_{3}+1}, \nonumber \\
w_{1}=w \gamma^{f_{1}-a/d}, & w_{2}=w \gamma^{f_{2}-(1-a/d)}, & w_{3}=w \gamma^{f_{3}-1}.
\end{eqnarray}
Coefficients $f_{1}$, $f_{2}$ and $f_{3}$ reflect the different values of activation energies for specific  polymerization and depolymerization processes. These coefficients  may be realistically  estimated as
\begin{eqnarray}
-a/d<f_{1}<a/d,\hspace{0.5cm} & -(1-a/d)<f_{2}<(1-a/d),\hspace{0.5cm} & -1<f_{3}<1
\end{eqnarray}    
It  means that the monomer attaches  faster to the place where the stronger lateral bond is created. Similarly, the dissociation is slower if larger lateral bond should be broken. 

For illustration purposes only, we consider  the case when $f_{1}=f_{2}=f_{3}=0$. The calculations for other possible values of parameters $f_{1}$, $f_{2}$ and $f_{3}$ produce qualitatively similar results. For the model with full dynamics we obtain  the following expression for the mean growth velocity,
\begin{eqnarray}
V=d\left(u-\frac{w}{\gamma}\right)\left(1+\frac{\gamma-1}{\gamma}\frac{\gamma^{a/d}}{1+\gamma^{2a/d-1}}\right).
\end{eqnarray}
In the ``one-layer'' model the velocity is given by
\begin{eqnarray}
V=d\left(u-\frac{w}{\gamma}\right)\frac{\gamma^{a/d}}{1+\gamma^{2a/d-1}}
\end{eqnarray}
To compare theoretical  predictions it is convenient to analyze the ratio of velocities, which gives us a measure of deviations between approximate and exact approaches. 

For the ratio of growth rates in two models we obtain 
\begin{equation}                                                                  
\frac{V_{one-layer}}{V_{N=2}}=\frac{1}{1+\gamma^{-a/d}+\gamma^{a/d-1}-\gamma^{-1}} \Rightarrow \left \{ 
\begin{array}{l}
 \frac{1}{2}, \gamma=1 \\  
  1,\gamma \gg 1 
\end{array} \right.                                                                     
\end{equation}
The ratio of velocities for two models is also plotted in Fig. 5 for different values of $a/d$. The simple analysis indicates that for any values of shift between the protofilaments the ratio of velocities  is approaching 1 for large $\gamma$. The convergence is better  with increasing values of $a/d$ and reaches the  maximum for $a/d=0.5$, i.e., when protofilaments are shifted by a half of the subunit length. But even for relatively small shifts, $a/d=0.1$, and realistic values of $\gamma \simeq e^{10}$ (energy of  lateral interaction between protofilaments is of order 10 $k_{B}T$), the deviation of the approximate ``one-layer'' model from full dynamics description is less than 10$\%$. Thus the ``one-layer'' model provides a very good approximation for the full dynamics description of the growth of rigid biopolymer with $N=2$ protofilaments.

\section{ Application of the One-layer Model for the Description of Experiments on Microtubule Growth}

The ``one-layer'' model can be easily used to describe the growth of real microtubules. To illustrate that we apply the approximate theory  to describe the experimental data of Dogterom and Yurke.\cite{dogterom97} In these experiments, the growing microtubules encountered a rigid microfabricated barrier, and the external forces have been calculated from buckling shapes of biopolymers for different growth velocities.

To describe this experimental force-velocity relation we used the following parameters: the rate constant  {\it u}  for the  process of association of a tubulin subunit to the tip of the leading protofilament  with a creation of the longitudinal (vertical) bond only, the  constant  {\it w} for the dissociation process from the leading protofilament which only breaks the longitudinal bond (see Fig. 3d for the biopolymer with $N=2$ protofilaments); the set of load distribution factors $\theta_{j}$, $1 \le j \le 13$; and the free energy of creation of lateral (horizontal) bond, $g_{h}$. All rate constants $u_{j}$ and $w_{j}$ can be expressed in terms of $u$, $w$, and the parameter $\gamma=\exp(-g_{h}/k_{B}T)$. To simplify the analysis, we also assumed that all load distribution factors are equal except $\theta_{1}$ and $\theta_{13}$. It reflects  the fact the external force may affect differently the rates of  removing or adding a subunit at the two adjacent protofilaments located at the seam. 

The growth velocity has been calculated using Eqs. (\ref{velocity}) and  (\ref{velocity.add}). The effect of external forces has been taken into account by using expressions (\ref{vel.force}). The rates $u_{j}$ and $w_{j}$ have been expressed in terms of rates $u$, $w$, and the parameter $\gamma$ in a way similar to $N=2$ case.  The resulting force-velocity curve is shown in Fig. 6. We find that the parameters
\begin{equation}
u=8.3 \mbox{ s}^{-1}, \quad w=355 \mbox{ s}^{-1}, \quad \theta_{j}=1 \mbox{ for all } j, \quad \gamma=100,
\end{equation}
provide a very satisfactory optimal fit to Dogterom and Yurke experimental data. Furthermore, this fit via the relation (\ref{stallforce}) allows us to predict the stall force $F_{S} \simeq 5.5$ pN.

As was explained above, our theoretical method also provides a connection with a phenomenological description. From the experimental fit we calculated the rates $k_{on}$ and $k_{off}$, which are observable rates of polymerization and depolymerization. Because the individual rates of association depend on concentration of free tubulin-GTP molecules, $u_{j}=k_{j}c$, at the limit of very large concentrations the mean growth velocity in the one-layer model is a linear function. This allows us to estimate $k_{on}$ and $k_{off}$ as follows,
\begin{equation}
k_{on}=N/ \sum_{j=1}^{N} k_{j}^{-1},  \mbox{  and   }  k_{off}=k_{on} c_{e},
\end{equation}
where $c_{e}$ is the critical concentration of tubulin monomers below which the biopolymer growth cannot happen. It can be easily calculated from Eq. (\ref{velocity}), $c_{e}=\prod_{j=1}^{N} w_{j}/k_{j}$. For experimental conditions of Dogterom and Yurke\cite{dogterom97} it is given by $c_{e} \simeq 10.6$ $\mu$M, which is only slightly larger than 2-5 $\mu$M, obtained in other experiments on microtubule assembly.\cite{vanden} Calculations from the experimental fit yields the following values for phenomenological rates,
\begin{equation}
k_{on}=3.0 \mbox{ s}^{-1}\mu\mbox{M}^{-1} \quad  \mbox{ and } \quad  k_{off}=31.8 \mbox{ s}^{-1}.
\end{equation}
 These estimates are in a good agreement with independent experimental observations, which put $k_{on}$ in the range of 2-10  s$^{-1}\mu\mbox{M}^{-1}$, while the experimental spread for $k_{off}$ is much larger, between 0.1 and 45 s$^{-1}$.

The ``one-layer'' model incorporates many microscopic properties of microtubules and it allows us to estimate some thermodynamic and structural properties of these biopolymers. The parameter $\gamma$ is associated with the lateral energy of interaction between the protofilaments. From the experimental fit we conclude  that this energy is given by $g_{h} \simeq -5 k_{B}T$, which is in excellent agreement with the only available estimates of -(3.2-5.7) $k_{B}T$, obtained from computer simulations of the stochastic model of microtubule assembly dynamics.\cite{vanburen02}. Note, however, that reasonable fits could also be produced with values of $\gamma$ ranging from 10 to 10$^{4}$.

Meanwhile the ratio of rates $u/w$ provides information on the energy of longitudinal bond in microtubules. At standard conditions (1 M solution of tubulin-GTP) this energy can be calculated from
\begin{equation}
g_{v}=-k_{B}T \ln \frac{u/c}{w} -g_{im},
\end{equation}
where $c=25$ $\mu$M is the concentration of free tubulins in experiments of Dogterom and Yurke.\cite{dogterom97} The parameter $g_{im}$ is a standard free energy of immobilization of a tubulin subunit in the polymer lattice, which was estimated to be in the range of 12-18 $k_{B}T$.\cite{erickson89}. Then, using the fitted values of rates $u$ and $w$ we obtain that $g_{v} \simeq -(19-25) \  k_{B}T$.  It is interesting to note that longitudinal interactions in microtubules  are much stronger than the lateral interactions between the protofilaments. These estimates are in excellent agreement with results of computer simulations of stochastic models of microtubule growth.\cite{vanburen02}

\section {Summary and Conclusions}

We constructed a simple stochastic microscopic model of the growth of rigid multi-filament biopolymers. It was argued that association/dissociation rates of individual monomers depend strongly on local environment, which leads to the conclusion that there is a finite number of polymer configurations that specify the dynamics of growing biopolymers. We suggested that most relevant configurations have protofilaments at  distances less than a monomer length from each other, i.e. ``one-layer'' configurations. As a result, the mean growth velocity and dispersion, or the effective diffusion constant, was calculated exactly for any number of protofilaments and for any shifts between them in terms of rate constants for attachments and detachments. It should be noted that simultaneous knowledge of the velocity and dispersion provides a  better description of fluctuations and variability in biopolymer dynamics. Precise experimental measurements of these properties would provide a valuable information on the growth mechanisms. 

Our approximate theoretical approach  easily takes into account the effect of external forces on growing biopolymers. It could be done by modifying rate constants using chemical-kinetic arguments. It allowed us then to construct a force-velocity relation and to estimate the stalling force, i.e., the force when the growth velocity becomes zero. It was suggested that the comparison with experimental force-velocity curves would provide a testing ground for this theoretical method.

Explicit expressions for the mean growth velocity and for dispersion obtained in the ``one-layer'' model allowed us to investigate the dependence of growth processes on monomers concentration. It is found that at large concentrations the mean growth velocity grows linearly, in agreement with phenomenological descriptions. However, at low concentrations the significant deviations from linearity may be found for some sets of parameters. This observation contradicts the main result of phenomenological models which assume that linear dependence of the growth velocity is valid at all concentrations. These non-linear deviations are probably due to the fact that the ``one-layer'' model provides a more realistic microscopic description of structural and geometrical properties of growing biopolymers, which apparently is more important at low concentrations of monomers. This question requires  careful experimental and theoretical tests. 

The validity and applicability of the ``one-layer''  model was discussed for the simple case of the growth of biopolymers with $N=2$ protofilaments. For this case an exact solution for full dynamics, that accounts for all possible polymer configurations,  was derived. It is found that the predictions of the approximate theory for the mean growth velocity are approaching the exact values for large, but realistic  values of lateral interactions between the protofilaments. This indicates that the ``one-layer'' model probably captures main physical and chemical properties of complex growth processes, and it can be used to describe  real systems, such as microtubules and actin filaments.

The advantages of using the ``one-layer'' model to describe the growth of rigid biopolymers is not only its simplicity and the ability to obtain explicit expressions for dynamic parameters, but also its very flexible and the fact that it can  be extended and modified in several directions in order to describe these complex processes more realistically. Firstly, more polymer configurations may be included by considering a ``two-layers'' condition, i.e., that the protofilaments in relevant configurations are at distances less than two monomer subunit lengths. More layers can be added in a similar fashion if required. Thus the results can be improved iteratively, i.e., this extension is analogous to a series expansion approach to the full dynamics description. Secondly, the apparently weaker interactions at the polymer lattice seam can be also incorporated. It will be interesting to know how this addition will effect the growth dynamics. Another possibility is to include tubulin-GTP hydrolysis in growth dynamics. This may give a new route to investigate the dynamic instability phenomena, which is still the most outstanding problem in the microtubule dynamics. Also, it will be interesting to compare the ``one-layer'' model with computer simulations of microtubule dynamics, which we plan to do in a future work. 

\section*{Acknowledgments}

We acknowledge the support from the Camille and Henry Dreyfus New Faculty Awards Program (under grant NF-00-056), the Welch Foundation (under grant no C-1559), and  US National Science Foundation through the grant CHE-0237105. We also thank M.E. Fisher  for critical discussions and encouragements.

\newpage

\noindent {\bf Figure Captions:} \\\\

\noindent Fig. 1 A typical configuration of growing rigid biopolymer consisting of $N$ protofilaments. The seam in the polymer lattice is between the protofilaments $N$ and 1. The tip of leading protofilament is at the origin. The free monomer can bind to any protofilament.    

\vspace{5mm}

\noindent Fig. 2 a) A biopolymer configuration with all protofilaments at distances less than one subunit length $d$ from the leading protofilament, i.e., a ``one-layer'' configuration. b) The biopolymer configuration which is not a ``one-layer'' configuration. The protofilament 2 is at the distance $a_{2} >d$.

\vspace{5mm}

\noindent Fig. 3 Four possible situations in the growth of a biopolymer with $N=2$ protofilaments.

\vspace{5mm}

\noindent Fig. 4 A kinetic scheme for  full dynamics description of the growth of a biopolymer with $N=2$ protofilaments.

\vspace{5mm}

\noindent Fig. 5 Ratio of mean growth velocities as a function of the parameter $\gamma$ for different protofilament shifts.

\vspace{5mm}

\noindent Fig. 6 The force-velocity curve from the experimental data of Dogterom and Yurke.\cite{dogterom97} The solid line represents the fit obtained using the ``one-layer'' model with load-dependent rate constants.

\newpage

\begin{figure}[ht]
\begin{center}
\vskip 1.5in
\unitlength 1in
\begin{picture}(3.4,2.8)
\resizebox{4.5in}{4.0in}{\includegraphics{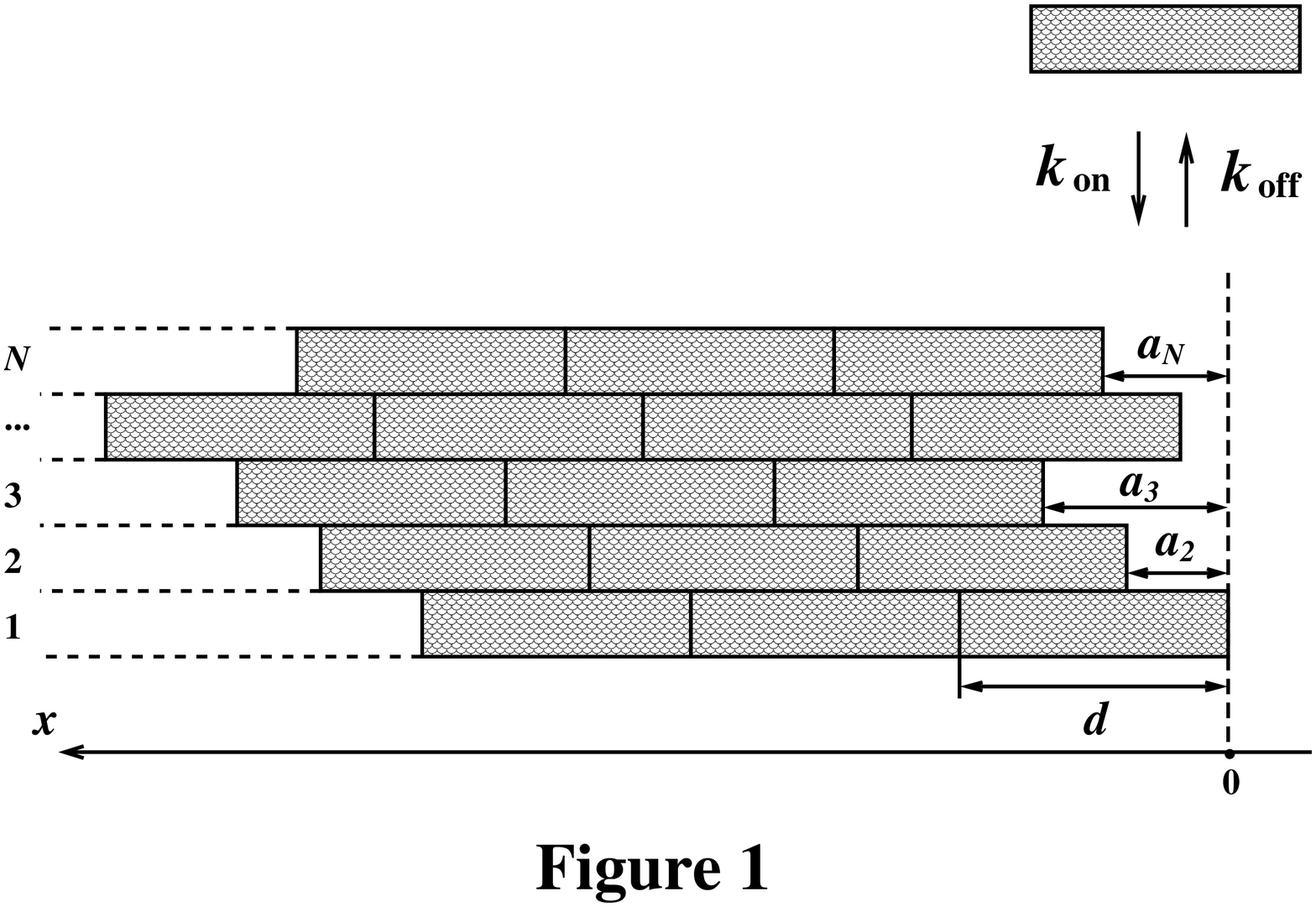}}
\end{picture}
\vskip 3in
 \begin{Large}  \end{Large}
\end{center}
\vskip 3in
\end{figure}

\begin{figure}[ht]
\begin{center}
\vskip 1.5in
\unitlength 1in
\begin{picture}(3.4,2.8)
\resizebox{4.2in}{4.5in}{\includegraphics{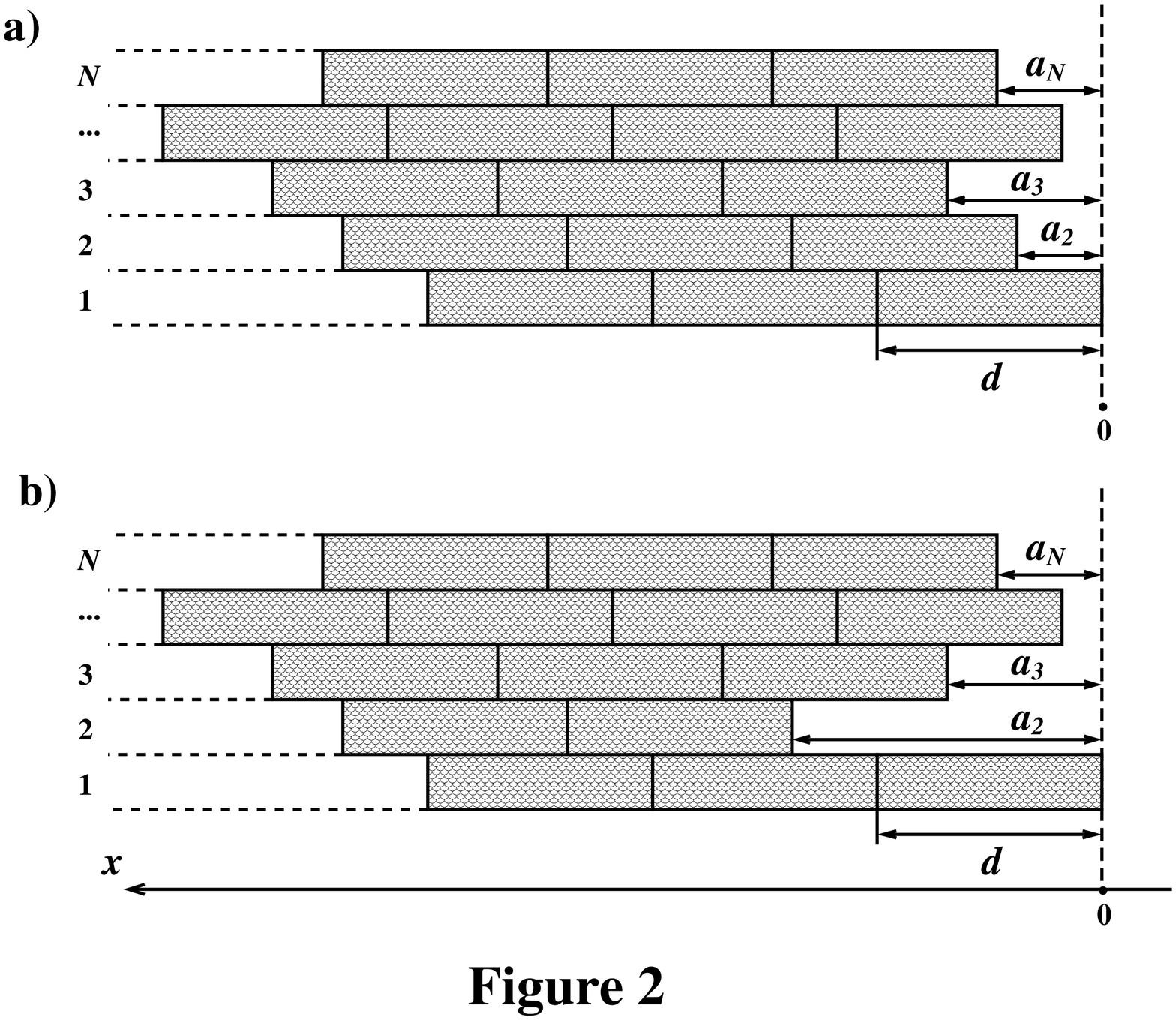}}
\end{picture}
\vskip 3in
 \begin{Large}  \end{Large}
\end{center}
\vskip 3in
\end{figure}

\begin{figure}[ht]
\begin{center}
\vskip 1.5in
\unitlength 1in
\begin{picture}(3.4,2.8)
\resizebox{4.5in}{4.7in}{\includegraphics{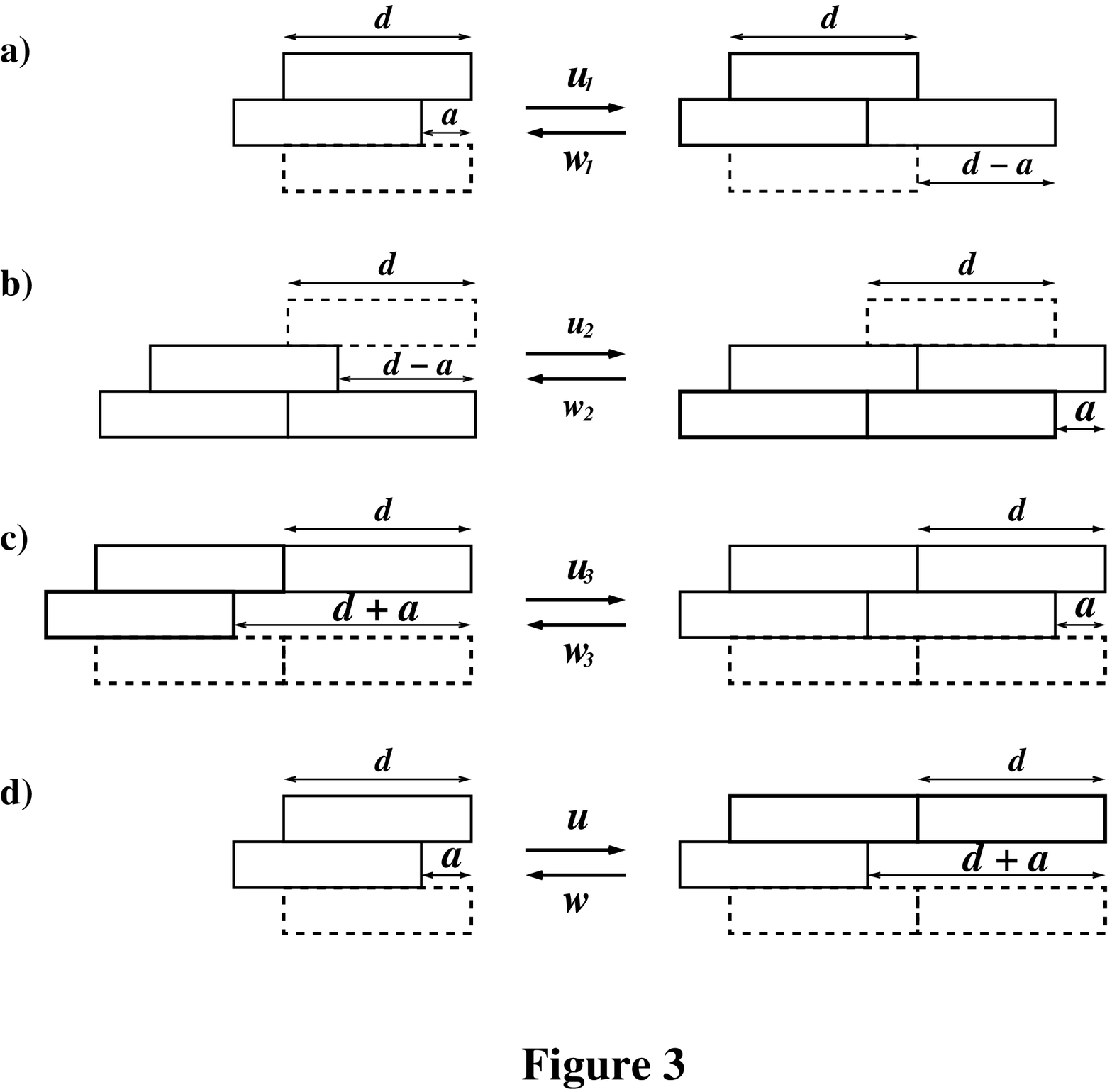}}
\end{picture}
\vskip 3in
 \begin{Large}  \end{Large}
\end{center}
\vskip 3in
\end{figure}

\begin{figure}[ht]
\begin{center}
\vskip 1.5in
\unitlength 1in
\begin{picture}(3.6,2.8)
\resizebox{5.0in}{4.0in}{\includegraphics{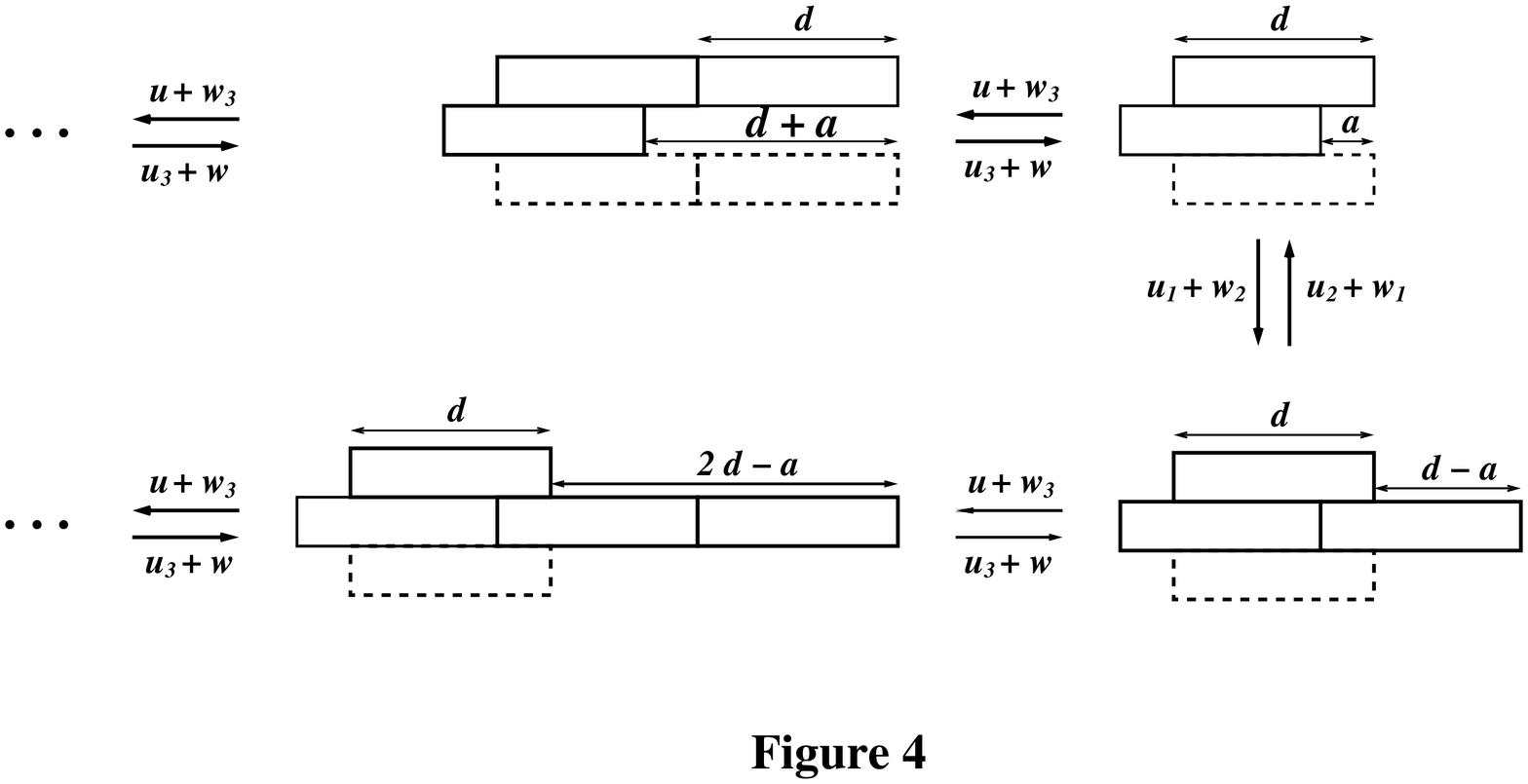}}
\end{picture}
\vskip 3in
 \begin{Large}  \end{Large}
\end{center}
\vskip 3in
\end{figure}

\begin{figure}[ht]
\begin{center}
\vskip 1.8in
\unitlength 1in
\begin{picture}(3.4,2.8)
\resizebox{5.0in}{4.0in}{\includegraphics{Figure5.eps}}
\end{picture}
\vskip 3in
 \begin{Large}  \end{Large}
\end{center}
\vskip 3in
\end{figure}

\begin{figure}[ht]
\begin{center}
\vskip 1.8in
\unitlength 1in
\begin{picture}(3.4,2.8)
\resizebox{5.0in}{4.0in}{\includegraphics{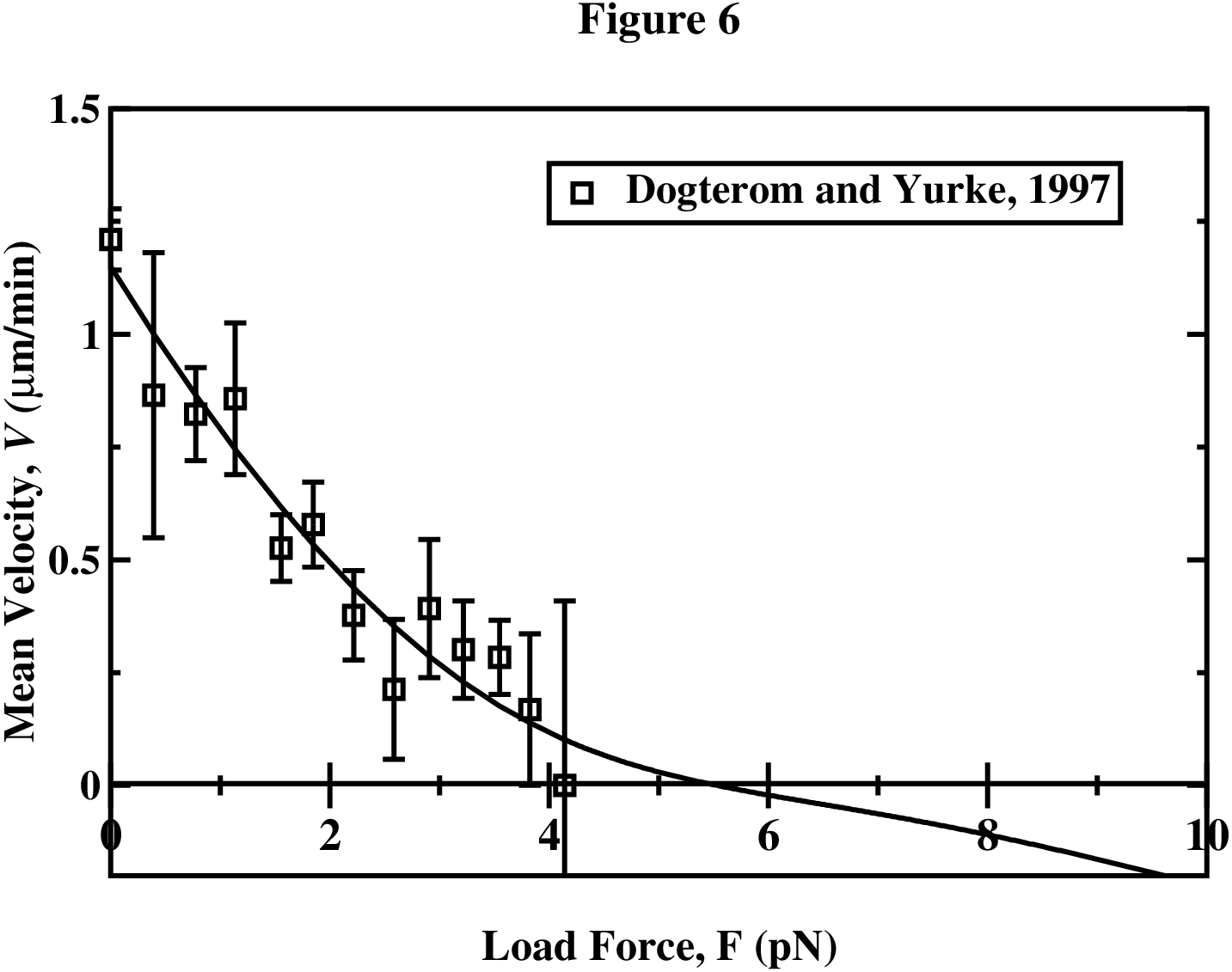}}
\end{picture}
\vskip 3in
 \begin{Large}  \end{Large}
\end{center}
\vskip 3in
\end{figure}

\end{document}